\documentclass[fleqn,twoside]{article}
\usepackage{espcrc2}

\usepackage{graphicx}
\usepackage[figuresright]{rotating}

\newcommand{\AmS}{{\protect\the\textfont2
  A\kern-.1667em\lower.5ex\hbox{M}\kern-.125emS}}

\title{Minimalist's Linux Cluster}

\author{Chang-Yeong Choi, Jeong-Hyun Kim, and Seyong Kim,
        \address[SJU]{Department of Physics, 
        Sejong University, Seoul 143-747, Korea}
        \thanks{Talk is presented by S.K. }}
       
\begin{document}

\begin{abstract}
Using barebone PC components and NIC's, we construct a linux cluster
which has 2-dimensional mesh structure. This cluster has smaller
footprint, is less expensive, and use less power compared to
conventional linux cluster. Here, we report our experience in building
such a machine and discuss our current lattice project on the machine.
\vspace{1pc}
\end{abstract}

\maketitle

\section{Motivation}

Constructing a Linux cluster using commodity PC's and commodity
networking hardware became quite easy and using such a cluster for a
lattice QCD project is very popular\cite{Thomas}. This increases the
range of computing power available to those who have only moderate
means. However, from our experience of using such a
cluster\cite{Lat99}, we found that there is a room for improvement in
scaling up the current Linux cluster architecture : first, if many
PC's are just stacked on top of each other in rows, soon the cluster
begins to occupy too large physical space. Secondly, not all the
components in a standard PC is essential for a lattice simulation. By
getting rid of unnecessary parts, one may reduce overall cost and
electrical power requirement for each PC's. Third, switched ethernet
hub is usually used in a linux cluster and providing full bisection
bandwidth using such switches is expensive. Building a cluster without
a switch may be more scalable. On the other hand, for those who have
limited resources like us, building everything (in particular,
hardware components) from the scratch to alleviate the above problems
is not sensible because it will probably take too long to develop such
components. Thus we looked for a solution which does not require
custom made hardware components and is re-usable in the future once
developed so that the evolutionary upgrade does not introduce delays.

\begin{figure}[t]
\begin{center}
\includegraphics[width=60mm,bb=161 250 452 552]{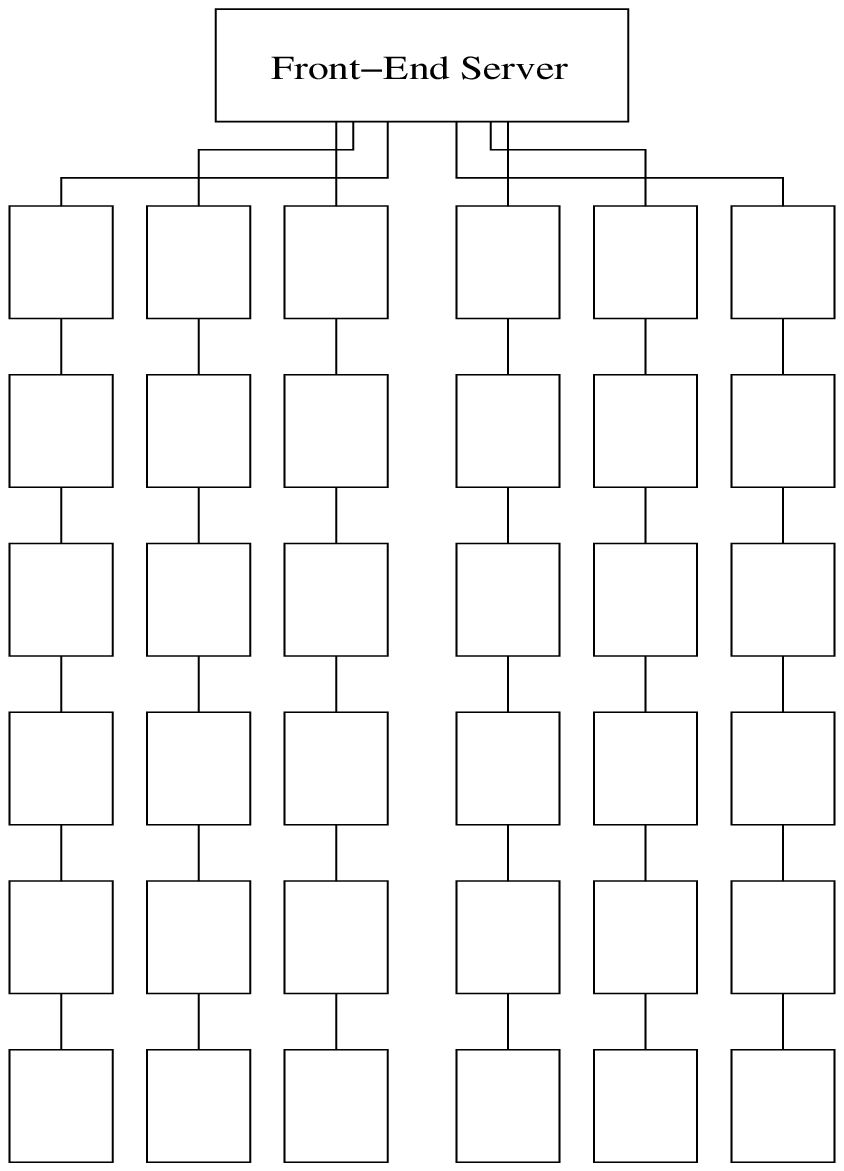}
\caption{Booting sequence}
\label{fig:booting}
\end{center}
\end{figure}

\section{Architecture and Hardware}

\begin{table*}[htb]
\caption{hardware components and prices}
\label{hardware}
\newcommand{\m}{\hphantom{$-$}}
\newcommand{\cc}[1]{\multicolumn{1}{c}{#1}}
\renewcommand{\tabcolsep}{1.4pc} 
\renewcommand{\arraystretch}{1.2} 
\begin{tabular}{@{}lrrr}
\hline
component         & unit price(in $\$$)  & no. of units & net price(in $\$$) \\
\hline
Intel P-IV 2.4GHz CPU    & 198  & 36  &  7,128 \\
ASUS P4-PE mother board  & 170  & 36  &  6,120 \\
512 MB PC2700 DDR SDRAM  & 93.5 & 36 & 3,366 \\
(3+1) RealTek 8139C NIC  & 44   & 36  & 1,584 \\
180W Sun ATX power supply   & 21   & 36 & 756  \\
network cable            & 8 & 36  & 288 \\
chasis($200\times 91\times 75$ cm)      & 1,037 & 1  & 1,037 \\
\hline
total price  &$534.5 \times 36 + 1,037 $ & & $= 20,279 $   \\
\hline
\end{tabular}\\[2pt]
\end{table*}

Each node is an extremely thin node which consists only of a Intel
pentium IV 2.4GHz CPU, 512 Mbytes DDR SDRAM, a mother board, and 4
fast ethernet network interface cards(NIC). One of 4 NIC's has a
socket for EPROM or EEPROM for the bootcode. Table\ref{hardware} shows
hardware components in the 36-node cluster and their costs. Except the
chasis and the network cables, everything is off-the-shelf components
and there is nothing special about them. The chasis is designed so
that each crates accept any standard ATX-size mother board and an
upgrade means just replacing mother boards with new one. The current
chasis size is suitable for 64-node configuration and has room for
additional 28 nodes. One PC with 360 GBytes hard disk serves as a
front end server.

Thus, development effort for our thin node cluster is mostly involved
with setting up necessary software environment : booting, OS, and MPI
parallel programming. Since there is no permanent storage device on
each nodes, booting is a little bit tricky and Linux operating system
needs to be configured dynamically after the boot. Fortunately, there
is a Linux solution, called ``Linux Terminal Server
Project''(LTSP)\cite{ltsp}, which is developed for the server-client 
situation similar to our case, a server booting up hosts of diskless
client computers. In this scenario, instead of booting from the kernel
image on a permanent media such as hard disk, floppy disk or flash
memory device, an NIC which has a small size EPROM or EEPROM (for
example 64 Kbytes) on the mother board does network booting. On
power-up, this network card on the client node executes its bootcode
and broadcast its IP request and its MAC address to the local network by
use of Dynamic Host Configuration Protocol(DHCP)\cite{dhcp}. The
server responds to this DHCP request and replies with the basic IP
information such as client node IP address, netmask setting, root file
directory and kernel image name depending on the client MAC addresses.
With the reply from the server, the client node configures its TCP/IP
and fetches kernel image from a host computer by Trivial File Transfer
Protocol(TFTP)\cite{net-howto}. Once the kernel image is loaded on the node
memory, the kernel starts executing and initializes the client node
and set it up for normal operation. One may choose whether application
softwares run on client nodes or runs on the server node.

The main difference between LTSP setup and ours lies on the assumed
network topology. LTSP relies on the star network connection and our
project adopts 2-dimensional mesh structure. In our case, each nodes
once booted, must act as a DHCP server and a TFTP server to the next
client node in constrast to the LTSP situation that central server
controls the other client nodes. This booting process may progress in
parallel to speed up : the front end server in our cluster starts
booting processes on 6 nodes simultaneously and then these 6 nodes
boot the next 6 nodes, etc (see Fig.\ref{fig:booting}). After the
booting procedure is completed, 2-D mesh routing is achieved by
explicit `route' command\cite{net-howto} in a script called from Linux
``init'' script. `route' assigns algorithmically one of four ethernet
devices, {\bf eth0}, {\bf eth1}, {\bf eth2}, and {\bf eth3} depending
on the destination IP addresses and the logical node ID. Since the
size of Linux routing table may grow upto 2048 elements by just
changing kernel compile option, this kind of explicit routing work
fine with a moderate size cluster. Ideally, one would like to have a
distributed routing mechanism implemented on the kernel level but it
is not part of the current Linux kernel. Linux distribution used on
the cluster is `Wow Linux version 7.1', which is equivalent to Red Hat
Linux 7.1 and the kernel version is 2.4.9. The version of LTSP package
which we modified for our need is 3.0.5. MPICH and LAM implementation
of MPI parallel programming environment is available on the cluster.

\section{Performance and Discussion}

\begin{figure}[t]
\begin{center}
\includegraphics[width=60mm,bb=161 250 452 552]{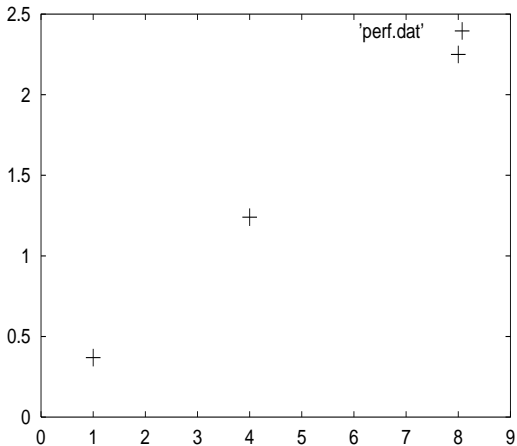}
\caption{Performance of the cluster. The horizontal axis is the number
  of nodes and the vertical axis is GFLOPS}
\label{fig:performance}
\end{center}
\end{figure}

Fig. \ref{fig:performance} shows the code performance of hybrid
molecular dynamics simulation of two staggered quark flavor with $m_q
a = 0.01$ on a $8^3 \times 512$ lattice (the single node benchmark is
for $8^3 \times 32$ lattice). One-dimensional ring ($N_t = 512$ is
distributed over the nodes) layout of lattice sites is used for the
code and the code is not yet optimal for the 2-D mesh structure of the
cluster. However, the code performance scales up nicely between 1 to 8
nodes. Sustained speed is about 2.25 GFLOPS on 8 node and is 11\% of
the theoretical peak speed. Thus, our cluster achieved $\sim 0.5$
MFLOPS/$\$$ with a straight FORTRAN code with no assembly language
subroutine. We find that using more than 8 nodes with the current full
QCD test code quickly degrades cluster performance due to non-optimal
communication pattern of the test code.

A conventional PC with Intel Pentium IV CPU is ordinarily equiped with
350W power supply. Since we put 180W power supply for each node and
the cluster operates fine with this condition, the overall power
requirement is successfully reduced to a half of usual Linux
cluster. Also, the footprint of our cluster is $200 \times 91 
\times 75$ cm, which is considerably smaller than that of stacking 36
PC's. The physical dimension of the full cluster (64 node) will be
even more beneficial since the same chasis will be used. Saved node
cost would be $\sim 100 \$ (\sim 15\%)$ from doing without a hard
disk. The whole construction is reusable as we planned since the
mother board size is the only factor which needs to considered in an
upgrade and the standard ATX size of mother board will stay with us
for a while.

Global MPI operations such as ``MPI\_ALLREDUCE'' involves multiple
hops in our cluster. Since LAM or MPICH relies on TCP/IP and each hops
contributes to software and hardware latency in message passing,
transversing many nodes reduces the efficiency of a program in our
cluster. However, since the software latency is larger than the hardware
latency, multiple hop will be less severe problem when user space
devices such as Infiniband\cite{infiniband} becomes cheaply available.


\begin{thebibliography}{9}
\bibitem{Thomas} Th. Lippert, plenary talk on ``Recent development of
  machines for lattice QCD'', in this proceeding.
\bibitem{Lat99} S. Kim, Nucl. Phys. {\bf B} (Proc. Suppl.),
  83-84(2000), 807.
\bibitem{ltsp} Linux Terminal Server Project, \\ http://www.ltsp.org.
\bibitem{dhcp} Dynamic Host Configuration Protocol, \\ http://www.dhcp.org.
\bibitem{net-howto} Linux Document Project, http://tldp.org.
\bibitem{infiniband} Infiniband architecture,\\
  http://www.infinibandta.org/home.
\end{thebibliography}
\end{document}